\documentclass[twocolumn,showpacs,preprintnumbers,amssymb,pra]{revtex4} 
\usepackage{graphicx}
\usepackage{dcolumn}
\usepackage{color}
\usepackage{bm}
\begin{document} 
%%%%%%%%%%%%%%%%%%%%%%%%%%%%%%%%%%%%%%%%%%%%%%%%%%%%%%%%%%%%%%%%%%%%%%%%%%%%%%%%
\newcommand{\ket}[1]{|#1\rangle}
\newcommand{\bra}[1]{\langle#1|}
\newcommand{\braket}[2]{\langle#1|#2\rangle}
\newcommand{\kb}[2]{|#1\rangle\langle#2|}
\newcommand{\kbs}[3]{|#1\rangle_{#3}\phantom{i}_{#3}\langle#2|}
\newcommand{\kets}[2]{|#1\rangle_{#2}}
\newcommand{\bras}[2]{\phantom{i}_{#2}\langle#1|}
\newcommand{\af}{\alpha}
\newcommand{\bt}{\beta}
\newcommand{\gm}{\gamma}
\newcommand{\la}{\lambda}
\newcommand{\dt}{\delta}
\newcommand{\s}{\sigma}
\newcommand{\qq}{(\s_{y}\otimes\s_{y})}
\newcommand{\uu}{\rho\qq\rho^{*}\qq}
\newcommand{\uuu}{\rho_{12}\qq\rho^{*}_{12}\qq}
\newcommand{\tr}{\textrm{Tr}} 
%%%%%%%%%%%%%%%%%%%%%%%%%%%%%%%%%%%%%%%%%%%%%%%%%%%%%%%%%%%%%%%%%%%%%%%%%%%%%%%%
%%%%%%%%%%%%%%%%%%%%%%%%%%%%%%%%%%%%%%%%%%%%%%%%%%%%%%%%%%%%%%%%%%%%%%%%%%%%%%%%
%\large

\title {\bf Control of atomic entanglement by dynamic Stark effect}
\author{Biplab Ghosh}
\altaffiliation{biplab@bose.res.in}
\affiliation{S. N. Bose National Centre for Basic Sciences,
Salt Lake, Kolkata 700 098, India}
\author{A. S. Majumdar}
\altaffiliation{archan@bose.res.in}
\affiliation{S. N. Bose National Centre for Basic Sciences,
Salt Lake, Kolkata 700 098, India}
\author{N. Nayak}
\altaffiliation{nayak@bose.res.in}
\affiliation{S. N. Bose National Centre for Basic Sciences,
Salt Lake, Kolkata 700 098, India}
\date{\today}

\vskip 0.5cm                              
\begin{abstract}
We study the entanglement properties of two three-level 
Rydberg atoms passing through a single-mode cavity. The interaction of 
an atom with the cavity field allows the atom to make a transition from 
the upper most (lower most) to the lower most (upper most) level by emission 
(absoprtion) of two photons via the middle level. We employ an effective 
Hamiltonian that describes the system with a Stark shifted two-photon 
atomic transition. We compute the entanglement of formation of the joint 
two-atom state as a function of Rabi angle $gt$. It is shown that the 
Stark shift can be used to enhance the magnitude of atomic 
entanglement over that obtained in the resonant condition for certain
parameter values. We find that though the two-atom entanglement generally
diminishes with the increase of the two-photon detuning and the 
Stark shift, it is 
possible to sustain the entanglement over a range of interaction times by
making the detuning and 
the Stark shift compensate each other. Similar characteristics are obtained
for a thermal state cavity field too.
\end{abstract}     
                                                            
\pacs{03.67.Mn,42.50.Hz}

\maketitle 

\section{Introduction}

The most interesting idea associated with composite quantum systems
is quantum entanglement. A pair of particles is said to be entangled in 
quantum mechanics if its state cannot be expressed as a product of the states 
of its individual constituents. Einstein, Podolsky 
and Rosen\cite{einstein} were the first to point out certain nontrivial
consequences of entanglement on the ontology of quantum theory. 
The preparation and manipulation of these 
entangled states lead to a 
better understanding of basic quantum phenomena. For example, complex 
entangled 
states, such as the Greenberger, Horne and Zeilinger\cite{greenberger} 
triplets of particles are used for tests of quantum nonlocality\cite{pan}. 
Beyond these fundamental aspects, entanglement has become a fundamental 
resource 
in quantum information processing, and there has been a rapid development 
of this subject in recent years\cite{zukowski}.

Cavity-QED has been a favourite tool to test the foundations of quantum 
mechanics including entanglement. Many beautiful experiments have been 
carried out, and in recent years, entangled states have been created and 
verified\cite{raimond}. Maximally entangled 
states between two modes in a single cavity have been generated 
using a Rydberg atom coherently 
interacting with each mode in turn\cite{rauschenbeutel}. Practical realization
of various features of quantum entanglement are obtained in atom-photon 
interactions in optical and microwave cavities.  Several 
studies have been performed to quantify the entanglement generated in 
atom-photon interactions in cavities\cite{masiak,kim,datta,ijqi2}.

The above cavity-QED related investigations involved mostly the 
absorption or emission 
of a single photon in an atomic transition. However, involvement of more than 
one photon, in particular, two photons in the transition between two atomic  
levels via a non-resonant intermediate level has been known for a long time
\cite{mcneil}. The output radiation from such interactions 
exhibits novel non-classical properties such as sub-Poissonian photon 
statistics. Needless to say, the idea of squeezed light has originated
from a two-photon process\cite{yuen}. Two-photon processes have 
also been studied in cavity-QED \cite{puri,bartzis,haroche,eberly}. It showed 
compact and regular quantum revivals in the atomic population in the single 
atom two-photon cavity-QED \cite{puri,bartzis}. Haroche and co-workers have 
demonstrated experimentally the
two-photon maser action in a micromaser cavity\cite{haroche}. In total,
the two-photon process mostly exhibits non-classical properties compared to 
the one-photon process\cite{jaynes,eberly}. Thus, it would be interesting to 
study the properties of atom-atom entanglement in the framework of a 
two-photon process. 

The two-photon atomic transition process also introduces a dynamic Stark 
shift in the atomic transition which is related to the magnitude of the
electric field of the radiation inside the cavity. This non-trivial effect 
which is naturally present in actual 
experiments involving two-photon transitions has to be properly acounted
for in  the its theoretical analysis, or example in the two-photon
micromaser\cite{haroche}. Various possibilities of exploiting the Stark effect
in quantum optical applications have been noticed in recent years. 
To name a few, schemes for applying the dc as well as the ac Stark shifts
towards implementation of quantum logic gates and algorithms\cite{biswas}, 
and in the 
improvement of photon
sources for interferometry\cite{milburn} have been suggested.
It is thus tempting to study if the Stark shift can be utilized to 
enhance atom-atom entanglement, as well. 

The purpose of the present 
paper is to investigate
the possibility of controlling atomic entanglement by the Stark shift
generated in atomic transitions inside cavities. 
In section II, we derive an effective Hamiltonian which efficiently
describes the atom-field two-photon interaction. This
Hamiltonian is quadratic in the field operators, which is at the root of 
nonclasssical properties that this process exhibits. The properties of
the atom-atom 
entanglement generated through this interaction is studied in section III.
We find that the atomic entanglement can be generally sustained, and for 
certain
interaction times enhanced too,  by making the
Stark shift compensate for the two-photon detuning. In section IV we show
that similar trends are also obtained for a thermal cavity field.
We conclude the paper with a summary of our results in section V.

\section{Derivation of the Effective Hamiltonian}

We consider a ladder system of a three-level Rydberg atom interacting with a 
single mode of a microwave cavity field. The middle level may be a group of 
closely spaced levels removed far away from one-photon resonance. Thus the 
interaction involves simultaneous absorption (or emission) of two photons
between the two atomic levels via a group of (or one) intermediate levels.
This is called a degenerate process since the two photons are from the same 
mode of the radiation field. Let us label the lower and the the upper level as 
$\ket{g}$ and $\ket{e}$ respectively and the intermediate levels are labelled 
as $\{\ket{i}\}$ (see Figure $1$). 

\vskip 0.02cm

\begin{figure}[h!]
\begin{center}
\includegraphics[width=6cm]{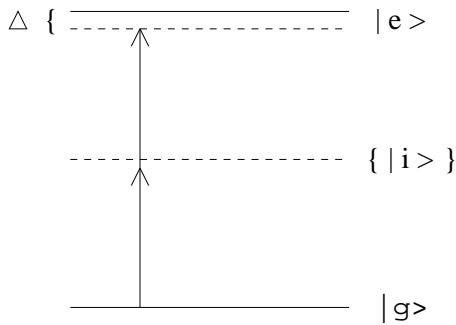}
\caption{A three-level Rydberg atom with its three energy 
levels are denoted by $\ket{e}$ (upper level), $\ket{i}$ (middle level),
$\ket{g}$ (lower level) respectively.}
\end{center}
\label{fc3}
\end{figure}

The atom initially in the lower state 
$\ket{g}$ absorbs a photon and jumps to one of the intermediate levels 
$\{\ket{i}\}$ and from there moves to the upper level $\ket{e}$ by absorbing
another photon. The return path of the atom from $\ket{e}$ to $\ket{g}$
is again via $\{\ket{i}\}$ by the emission of two photons to the same mode.
To make the equations look simpler let us consider only one intermediate 
level $\ket{i}$. A microscopically correct Hamiltonian describing the above 
process can be written as 
\begin{eqnarray}
H=H_0+H_1
\end{eqnarray}
where the interaction Hamiltonian $H_1$ is of the order of one-photon dipole 
interaction strength. The $H_0$ and $H_1$ can be written in operator form  
as  
\begin{eqnarray}
H_0=\omega_e\kb{e}{e}+\omega_i\kb{i}{i}+\omega_g\kb{g}{g}+\omega a^\dagger a
\end{eqnarray}
and 
\begin{eqnarray}
H_1=g_1(S^{+}_{gi}a+S^{-}_{gi}a^\dagger)+g_2(S^{+}_{ei}a+S^{-}_{ei}a^\dagger)
\end{eqnarray}
respectively. $g_1$ and $g_2$ are coupling constants (in dipole approximation) 
for the one-photon interactions responsible for the transitions between 
the level $\ket{g}$ and $\ket{i}$ and that between $\ket{e}$ and $\ket{i}$
respectively. The atomic operators are given by $S^{+}_{gi}=\kb{i}{g}$
and $S^{-}_{gi}=\kb{g}{i}$ and similar definitions go for the operators 
involving the upper level $\ket{e}$. In the basis of the states 
$\ket{e,n}$, $\ket{i,n+1}$ and $\ket{g,n+2}$, allowed by the rotating-wave
approximation, we can write the Hamiltonian in the form of a $3\times3$
matrix. The density matrix of the atom-field system obeys the equation
\begin{eqnarray}
i\dot\rho=[H,\rho].
\end{eqnarray}
The equation of motion can be solved by one of the usual known methods, but, 
the derivations get more and more tedious as the number of 
intermediate levels increases. However, as these levels are removed far away 
from one-photon resonance, we can use this to obtain an effective Hamiltonian 
which efficiently describes the two-photon process. In doing so, we follow 
the method outlined in Refs.\cite{bartzis,takatsuji}.
We start by making a canonical transformation
\cite{bartzis,takatsuji} $\rho_r=e^{-iK}\rho e^{iK}$
where $K$ is time-independent and Hermitian. $\rho_r$ obeys the equation 
of motion 
\begin{eqnarray}
i\dot\rho_r=[H_{eff},\rho_r]
\end{eqnarray}
where 
\begin{eqnarray}
H_{eff}&=&e^{-iK}He^{iK}\nonumber\\
&=&H-i[K,H]-1/2[K,[K,H]]+........
\label{ec2}
\end{eqnarray}
We know that the probability of one-photon transition is inversly 
proportional to the one photon detuning $\omega_{ki}-\omega$ 
($k\equiv e,g$). Since this detuning is large, the one-photon transition
probabilities are very small. In this situation, it is safe to retain terms
upto second order in one-photon coupling constants. 
Then, retaining terms of the order of square of the 
coupling constants in the above expression, we have 
\begin{eqnarray}
H_{eff}=H_0+H_1-i[K,H_1]-\frac{1}{2}[K,[K,H_0]].
\label{ec3}
\end{eqnarray}
Since $K$ is arbitrary, we choose that
\begin{eqnarray}
[K,H_0]=-iH_1,
\label{ec4}
\end{eqnarray}
from which we determine all the elements of $K$. This reduces the effective 
Hamiltonian to
\begin{eqnarray}
H_{eff}=H_0-i/2[K,H_1].
\label{ec5}
\end{eqnarray}
Written back in operator form, $H_{eff}$ takes the form 
\begin{eqnarray}
H_{eff}&=&[\Delta+(\beta_e+\beta_g)a^\dagger a]S_z +\frac{1}{2}(\beta_e-\beta_g)a^\dagger a\nonumber\\
&+&G(S^+a^2+S^-{a^\dagger}^2)
\label{ec6}
\end{eqnarray}
where the spin operators are $S_z=\frac{\kb{e}{e}-\kb{g}{g}}{2}$, 
$S^+=\kb{e}{g}$
and $S^-=\kb{g}{e}$.
$\Delta$ is the two-photon detuning and is given by 
$\Delta=\omega_e-\omega_g-2\omega$ and $G$ is the two-photon coupling constant 
having the form  
\begin{eqnarray}
G=\sum_i\frac{g_1g_2}{2}\sqrt{(n+1)(n+2)}[\frac{1}{\omega_{ei}-\omega}-
\frac{1}{\omega_{ig}-\omega}]
\end{eqnarray}
where $\omega_{ei}=\omega_{e}-\omega_{i}$ and 
$\omega_{ig}=\omega_{i}-\omega_{g}$. The Stark shifts associated 
with the levels 
$e$ and $g$ are, respectively, 
\begin{eqnarray}
\beta_e=\sum_i\frac{{g_2}^2}{\omega_{ei}-\omega}
\end{eqnarray}
and 
\begin{eqnarray}
\beta_g=\sum_i\frac{{g_1}^2}{\omega_{ig}-\omega}.
\end{eqnarray}

We now have a Hamiltonian describing the interaction between an an effective 
two-level system or an effective spin-$1/2$ system and a single mode radiation 
field of frequency $\omega$. If we take $\beta_e=\beta_g=\beta$ the Hamiltonian reduces to
\begin{eqnarray}
H_{eff}=[\Delta+2\beta a^\dagger a]S_z+G(S^+a^2+S^-{a^\dagger}^2)
\label{ec7}
\end{eqnarray}
It may be noted here that $H_{eff}$ is a function of the two-photon detuning 
$\Delta$ which is an outcome of the procedure followed here. In other words, 
we need not assume the resonance condition $\Delta=0$ unlike in other methods 
in literature\cite{puri}, but this method gives the two-photon detuning 
$\Delta$ as an independent parameter for the analysis.
The effective Hamiltonian can easily be contrasted when compared with a
Hamiltonian describing one-photon process \cite{jaynes}. First, the effective
Hamiltonian is now cavity photon number dependent and is thus dynamic.
Secondly, the Hamiltonian is quadratic in annihilation and creation operators.
As mentioned earlier, they are at the root of all the nonclassical behaviours 
in two-photon processes. Using such a Hamiltonian, we now study the 
entanglement of two such atoms passing through the cavity one after the other.

\section{Two-atom entanglement}

The effective Hamiltonian derived above 
can be written in the matrix form in the basis
of $|e,n>$, $|g,n+2>$ states as
\begin{eqnarray}
H_{eff}=\left(\begin{matrix}{(\frac{\Delta}{2}+\beta n)&g \sqrt{(n+1)(n+2)} \cr
g \sqrt{(n+1)(n+2)}&-(\frac{\Delta}{2}+\beta n+2\beta)}\end{matrix}
\right).
\label{ec8}
\end{eqnarray}
The eigenvalues of $H_{eff}$ are 
\begin{eqnarray}
\la_1=-\beta+\sqrt{[\frac{\Delta}{2}+\beta (n+1)]^2+g^2(n+1)(n+2)},\\
\la_2=-\beta-\sqrt{[\frac{\Delta}{2}+\beta (n+1)]^2+g^2(n+1)(n+2)}.
\end{eqnarray}
The corresponding eigenstates can be written as 
\begin{eqnarray}
\kets{\Psi}{\la_1}=c_1\ket{e,n}+c_2\ket{g,n+2},\\
\kets{\Psi}{\la_2}=c_2\ket{e,n}-c_1\ket{g,n+2},
\end{eqnarray}
where 
\begin{eqnarray}
c_1=\frac{\la_1+(\frac{\Delta}{2}+\beta n+2\beta)}{\sqrt{g^2(n+1)(n+2)+(\la_1+(\frac{\Delta}{2}+\beta n+2\beta))^2}}
\end{eqnarray}
and
\begin{eqnarray}
c_2=\frac{g \sqrt{(n+1)(n+2)}}{\sqrt{g^2(n+1)(n+2)+(\la_1+(\frac{\Delta}{2}+\beta n+2\beta))^2}}.
\end{eqnarray}

We envisage a process in which two atoms pass through an ideal cavity 
($Q=\infty$) such that there is no overlap of their flights there 
\cite{datta,ijqi2}. The interaction of each atom in the 
cavity is described by $H_{eff}$ in Eq.(\ref{ec8}). We assume that the two 
atoms are in their respective upper states $\ket{e}$ before they enter the 
cavity empty of photons. After passage of the first atom through the cavity, 
the joint atom-field state is given 
at any time $t$ is
\begin{eqnarray}
\kets{\Psi_1}{t}=(r_1-is_1)\ket{e,n}+(r_2-is_2)\ket{g,n+2}.
\end{eqnarray}
where
\begin{eqnarray}
r_1&=&[c_1^2\cos{(\la_1t)}+c_2^2\cos{(\la_2t)}],\\
s_1&=&[c_1^2\sin{(\la_1t)}+c_2^2\sin{(\la_2t)}],\\
r_2&=&c_1c_2[\cos{(\la_1t)}-\cos{(\la_2t)}],\\
s_2&=&c_1c_2[\sin{(\la_1t)}-\sin{(\la_2t)}].
\end{eqnarray}
The second atom then interacts with the cavity field modified 
by the passage of the first 
atom. Assuming the flight time of the two atoms 
through the cavity to be the same, 
the joint state of the two atoms and the cavity after the second leaves the 
cavity is given by
\begin{eqnarray} 
\kets{\Psi_{12}}{t}&=&(r_1-is_1)^2\ket{e_1,e_2,n}\nonumber\\
&+&(r_1-is_1)(r_2-is_2)\ket{e_1,g_2,n+2}\nonumber\\
&+&(r_2-is_2)(r_1^{\prime}-is_1^{\prime})\ket{g_1,e_2,n+2}\nonumber\\
&+&(r_2-is_2)(r_2^{\prime}-is_2^{\prime})\ket{g_1,g_2,n+4},
\end{eqnarray}
where $r_1^{\prime}=r_1^{n=n+2}$, $s_1^{\prime}=s_1^{n=n+2}$, 
$r_2^{\prime}=r_2^{n=n+2}$ and $s_2^{\prime}=s_2^{n=n+2}$.
Next we calculate the two-atom mixed state taking trace over the field 
variables. The joint two-atom mixed state density matrix  in the basis
of $|e_1,e_2>$, $|e_1,g_2>$, $|g_1,e_2>$ and  $|g_1,g_2>$ states is given by
\begin{eqnarray}
\rho_{12}=\left(\begin{matrix}{\alpha&0&0&0 \cr
0&\gamma&\epsilon&0\cr0&\epsilon^*&\delta&0\cr0&0&0&\eta}\end{matrix}
\right).
\label{matrix1}
\end{eqnarray} 
where $$\alpha=(r_1^2+s_1^2)^2,$$ $$\gamma=(r_1^2+s_1^2)(r_2^2+s_2^2),$$
$$\delta=(r_2^2+s_2^2)({r_1^{\prime}}^2+{s_1^{\prime}}^2),$$
$$\eta=(r_2^2+s_2^2)({r_2^{\prime}}^2+{s_2^{\prime}}^2),$$ and 
$$\epsilon=(r_2^2+s_2^2)(r_1-is_1)(r_1^{\prime}+is_1^{\prime}).$$

We compute the two-atom entanglement using the well-known measure of the
entanglement of formation\cite{hill} given by  
\begin{eqnarray}
E_{F}(\rho)=h\left(\frac{1+\sqrt{1-C^{2}(\rho)}}{2}\right),
\end{eqnarray}
where $C$ is called the concurrence defined by the formula
\begin{eqnarray}
C(\rho)=\max(0, \sqrt\la_1-\sqrt\la_2-\sqrt\la_3-\sqrt\la_4)
\end{eqnarray}
where the 
$\la_{i}$ are the  eigenvalues of $\uu$ in descending order,
and 
\begin{eqnarray}
h(x)=-x\log_{2}x-(1-x)\log_{2}(1-x)
\end{eqnarray}    
is the binary entropy function.
The entanglement of formation is monotone of concurrence. The 
eigenvalues of $\uuu$ in this case are given by
$\alpha\beta$, $\alpha\beta$, $(\sqrt{\gamma\delta}+|\epsilon|)^2$ and
$(\sqrt{\gamma\delta}-|\epsilon|)^2$ respectively.

\vskip 1cm
\begin{figure}[h!]
\begin{center}
\includegraphics[width=6.5cm]{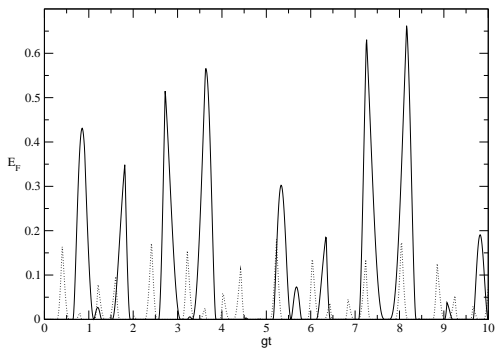}
\caption{$E_F$ is plotted vs Rabi angle $gt$ for (i) $\Delta/g=0$ and 
$\beta/g=0$ (solid line) (ii) $\Delta/g=2$ and $\beta/g=2$ (dotted line).}
\end{center}
\end{figure}

We compute numerically 
the entanglement of formation $E_F$ for the two atoms, and plot it versus the
Rabi angle $gt$ for different combinations of the two-photon detuning
$\Delta$ and the Stark shift $\beta$ in the Figures 2 and 3. 
We find that entanglement between the two 
atoms is controlled by the two-photon detuning and the Stark-shift parameters.
We first plot $E_F$ in Fig.2 for the resonant condition ($\Delta=0$ and
$\beta=0$). Now, if one includes a non-vanishing detuning and the resultant
Stark shift, one sees from Fig.2 that the two-atom entanglement is diminished 
in general over a range of values of the Rabi angle. It can be verified that
as long as  $\Delta$ and $\beta$ are of the same sign, 
the magnitude of entanglement between the two atoms decreases with the 
increase of $\Delta$ or $\beta$.

\vskip 1cm
\begin{figure}[h!]
\begin{center}
\includegraphics[width=6.5cm]{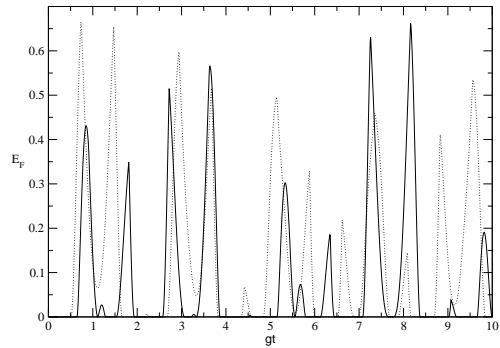}
\caption{$E_F$ is plotted vs Rabi angle $gt$ for (i) $\Delta/g=0$ and 
$\beta/g=0$ (solid line) (ii) $\Delta/g=-1$ and $\beta/g=1$ (dotted line).}
\end{center}
\end{figure}

The more interesting case is obtained when the detuning and the Stark shifts
are of opposite signs. This is revealed in Fig.3 where the two-atom 
entanglement $E_F$ versus $gt$ for $\Delta=\beta=0$ is compared with 
$E_F$ when $\beta=-\Delta=1$. We see that the magnitude of the 
entanglement between the two atoms increases when $\beta=-\Delta=1$ with 
respect to the case when $\Delta=\beta=0$ for certain values of the 
Rabi angle, e.g., $gt \approx 1$. Such enhancement of entanglement is
observed in varying measures for other values of the Rabi angle also.
Its origin lies in the presence of the photon number operator $a^\dagger a$ 
in the effective Hamiltonian in Eq.(14) which makes the ``effective'' 
two-photon detuning 
$\Delta +2\beta a^\dagger a$ dynamic as the cavity photon 
number oscillates in time. 
One sees over a range of interaction times that the atomic entanglement
is in general sustained with variations in oscillatory behaviour with 
respect to $gt$, if the two-photon detuning $\Delta$ is 
compensated by the Stark shift.
Or in other words, the maximum entanglement that can be obtained by
varying $gt$ over a range of interaction times remains similar to
that in the resonant case.
This is in striking contrast to the case displayed in Fig.2 where the 
entanglement reduces substantially with the increase of $\Delta$ and 
$\beta$, both having the same sign. It is interesting to note that the 
entanglement between two atoms is also preserved if we interchange the 
sign of $\Delta$ and $\beta$ but with the condition $\Delta+\beta=0$.
Overall, we find that the Stark shift 
acts as a control parameter for the atom-atom entanglement.

\section{Atomic entanglement mediated by the thermal field}

The thermal field is the most easily available radiation field, and so, 
its influence on the 
entanglement of spins is of much interest. The atomic entanglement 
mediated by the thermal field through single photon processes have
been studied earlier\cite{ijqi2,kim}. Since the thermal field is related
to the temperature of the medium, photons of this field are naturally 
present inside the cavity. So it is not out of place to include the Bose 
statistics for the thermal field in our analysis. The field at thermal 
equilibrium obeying Bose-Einstein statistics has an average photon number 
at temperature $T^0 K$, given by 
\begin{eqnarray}
<n>=\frac{1}{e^{\hbar \omega/kT}-1}.
\label{24}
\end{eqnarray}
The photon statistics is governed by the distribution $P_n$ given by
\begin{eqnarray}
P_n=\frac{<n>^n}{(1+<n>)^{n+1}}.
\label{25}
\end{eqnarray}
This distribution function always peaks at zero, i.e., $n_{peak}=0$.
For a field in a thermal state, the joint  
two-atom-cavity state is obtained by summing over all $n$, and is given by 
\begin{eqnarray} 
\kets{\Psi_{12}}{t}&=&\sum_nA_n[(r_1-is_1)^2\ket{e_1,e_2,n}\nonumber\\
&+&(r_1-is_1)(r_2-is_2)\ket{e_1,g_2,n+2}\nonumber\\
&+&(r_2-is_2)(r_1^{\prime}-is_1^{\prime})\ket{g_1,e_2,n+2}\nonumber\\
&+&(r_2-is_2)(r_2^{\prime}-is_2^{\prime})\ket{g_1,g_2,n+4}],
\end{eqnarray}
where $P_n=|A_n|^2$ is the photon distribution function of the thermal field.
The reduced mixed density matrix of two atoms after passing through the 
the thermal cavity field in the basis
of $|e_1,e_2>$, $|e_1,g_2>$, $|g_1,e_2>$ and  $|g_1,g_2>$ states is given by
\begin{eqnarray}
\rho_{12}=\left(\begin{matrix}{\alpha_1&0&0&0 \cr
0&\gamma_1&\epsilon_1&0\cr0&\epsilon_1^*&\delta_1&0\cr0&0&0&\eta_1}\end{matrix}
\right).
\label{matrix2}
\end{eqnarray} 
where $$\alpha_1=\sum_nP_n(r_1^2+s_1^2)^2,$$ $$\gamma_1=\sum_nP_n(r_1^2+s_1^2)
(r_2^2+s_2^2),$$
$$\delta_1=\sum_nP_n(r_2^2+s_2^2)({r_1^{\prime}}^2+{s_1^{\prime}}^2),$$
$$\eta_1=\sum_nP_n(r_2^2+s_2^2)({r_2^{\prime}}^2+{s_2^{\prime}}^2)$$ and 
$$\epsilon_1=\sum_nP_n(r_2^2+s_2^2)(r_1-is_1)(r_1^{\prime}+is_1^{\prime}).$$
 
\vskip 1cm
\begin{figure}[h!]
\begin{center}
\includegraphics[width=6.5cm]{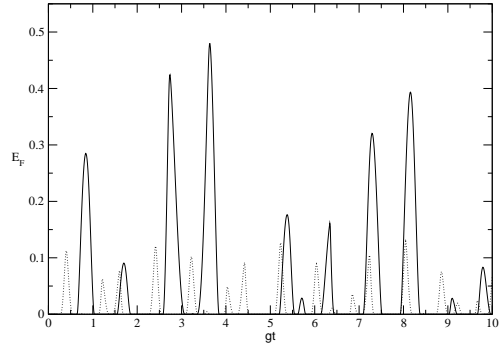}
\caption{$E_F$ is plotted vs Rabi angle $gt$ for (i) $\Delta/g=0$ and 
$\beta/g=0$ (solid line) (ii) $\Delta/g=2$ and $\beta/g=2$ (dotted line). The 
average thermal photon number $<n>=0.1$.}
\end{center}
\end{figure}

We again compute the entanglement of formation of the joint two-atom
state after it emerges from the cavity. Similar to the case of the vacuum
cavity field considered in Section III, the thermal field mediates
entanglement between the two atoms even though there is no direct
interaction between them. This feature was also observed earlier in
context of the one-photon atomic transition process\cite{kim,ijqi2}.
The variation  of the magnitude of the two-photon entanglement versus
the Rabi angle is displayed in the Figs. 4, 5 and 6. 

\vskip 1cm 
\begin{figure}[h!]
\begin{center}
\includegraphics[width=6.5cm]{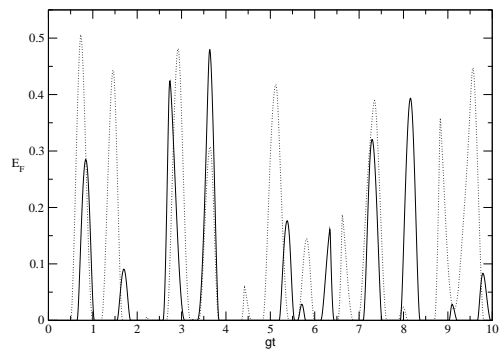}
\caption{$E_F$ is plotted vs Rabi angle $gt$ for (i) $\Delta/g=0$ and 
$\beta/g=0$ (solid line) (ii) $\Delta/g=-1$ and $\beta/g=1$ (dotted line).
The average thermal photon number $<n>=0.1$.}
\end{center}
\end{figure}

Our purpose here is to investigate the effect of Stark shift on atomic
entanglement, and to this end 
we plot in Figs. 4 and 5 the entanglement of formation as a function of 
the Rabi angle $gt$ for different combinations of the two-photon detuning 
$\Delta$ and the Stark shift parameter $\beta$ when the thermal field 
has an average photon number $<n>=0.1$.
We notice that the variation in $E_F$ as a function of $gt$ is similar 
(barring differences in magnitudes) to the case of vacuum cavity field.
From Fig.4, we note again, that as long as  $\Delta$ and $\beta$ are of 
same sign, the magnitude of entanglement between the two atoms decreases 
compared to the resonant case with the increase of $\Delta$ or $\beta$.
But, as seen from Fig.5, atomic entanglement can be increased for
particular values of $gt$ by choosing $\beta$ to 
be of opposite sign as $\Delta$.
Again, similar to the case of the vacuum cavity field, we find that 
if the two-photon detuning is compensated by the Stark shift (Fig. $5$),
the atomic entanglement can be sustained on average over a range of
values of the Rabi angle. These characteristics are still noticed for 
higher values of the average thermal photon number $<n>$. As seen from
Fig.6, only the magnitude
of $E_F$ is reduced with increase of average thermal photons. 
Thus, the Stark shift can be used to control
the atomic entanglement mediated by the thermal field
by preserving the maximum magnitude of
entanglement obtained in a large range of atom-cavity interaction times.

\vskip 1cm 

\begin{figure}[h!]
\begin{center}
\includegraphics[width=6.5cm]{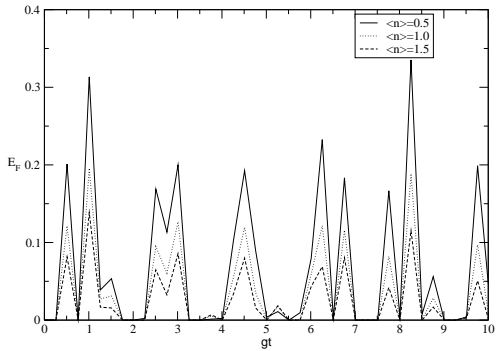}
\caption{$E_F$ is plotted vs Rabi angle $gt$ for $\Delta/g=-2$ and 
$\beta/g=2$.}
\end{center}
\end{figure}

\section{Conclusions} 

In this paper we have investigated the possibility of the control of
atomic entanglement by the Stark shift
generated in atomic transitions inside cavities. To this end we have
considered a degenerate two-photon process in a ladder 
system and obtained an effective Hamiltonian which describes the 
interaction efficiently\cite{bartzis,takatsuji}. The two-atom entanglement 
is shown to be 
mediated by the cavity field through which the two atoms pass successively
without any spatial overlap between them\cite{datta,ijqi2}. 
We are able to use the two-photon detuning
which comes out naturally in the method presented in Section II, to 
compensate the dynamic Stark
shift to get the atom-atom entanglement. Through this method we have
shown the possibility of using the dynamical Stark shift in controlling 
atomic entanglement mediated both by the vacuum cavity field as well
as by the thermal cavity field.

We have shown that the entanglement between two atoms 
depends on the 
two-photon detuning and the Stark shift parameter. The magnitude of atomic
entanglement quantified by the entanglement of formation diminishes
with the increase of the detuning and the stark shift. However, interestingly,
we have found that such a trend could be reversed if the values of the detuning
and the Stark shift are made to compensate each other. In the latter case
the entanglement could be even enhanced compared to the resonant situation
for particular values of the atom-photon interaction time. More generally,
it has been shown that the maximum magnitude of entanglement generated
over a range of values of the Rabi angle is nearly sustained if 
we set the values of the two-photon detuning and the stark-shift to be equal 
and opposite in sign. The effects of photon statistics of the thermal
field on the mediated entanglement\cite{kim,ijqi2} has also been studied. 
We have shown
that the general characteristics of the atomic entanglement as a 
function of the photon detuning and the stark shift parameter are
maintained for the case of the thermal field inside the cavity. 
The use of Stark shifts in some quantum information protocols has
been suggested recently\cite{biswas,milburn}, and an experiment to demonstrate
the enhancement of Rydberg atom interactions has actually been 
perfomed\cite{bohlouli}. Our present study should motivate further 
investigations on the feasability of using Stark shifts in the
practical manipulation of quantum entanglement.

\end{document}